\documentstyle[twoside,fleqn,espcrc2,epsf]{article}

\newcommand\as{\alpha_{\rm{S}}}

\def\beq{\begin{equation}}
\def\eeq{\end{equation}}
\def\bea{\begin{eqnarray}}
\def\eea{\end{eqnarray}}

\def\np#1#2#3{Nucl.\ Phys.\ B#1 (19#3) #2}
\def\pl#1#2#3{Phys.\ Lett.\ B#1 (19#3) #2}

\def\zp#1#2#3{Zeit.\ Phys.\ C#1 (19#3) #2}

\title{Quark mass effects in QCD jets
\thanks{Supported by the Conselleria de Cultura, Educaci\`o i 
Ci\`encia de la Generalitat Valenciana and CICYT, Spain, 
under grant AEN-96/1718.}}
\author{G.~Rodrigo
\address{Dept. de F\'{\i}sica Te\`orica,
Univ. de Val\`encia, \\ E-46100 Burjassot, Valencia, Spain.
E-mail: rodrigo@titan.ific.uv.es}}

\begin{document}

\begin{titlepage}
\renewcommand{\thefootnote}{\fnsymbol{footnote}}
\begin{flushright}
     hep-ph/xxxxxxx \\ FTUV/96-44 \\ IFIC/96-52
     \end{flushright}
\par \vspace{10mm}
\begin{center}
{\Large \bf
Quark mass effects in QCD jets
\footnote{to be published in the Proceedings of the High 
Energy Physics International Euroconference on Quantum 
Chromodinamics ({\it QCD '96}), Montpellier, France, July 1996.
Ed. S. Narison, Nucl Phys. B (Proc. Suppl.).}}
\end{center}
\par \vspace{2mm}
\begin{center}

{\bf G.~Rodrigo}\\

\vspace{5mm}

{Dept. de F\'{\i}sica Te\`orica, Univ. de Val\`encia,} \\
{E-46100 Burjassot, Valencia, Spain \\
E-mail: rodrigo@titan.ific.uv.es}

\end{center}

\par \vspace{2mm}
\begin{center} {\large \bf Abstract} \end{center}
\begin{quote}
We present the calculation of the decay width of the $Z$-boson 
into three jets including complete quark mass effects to 
second order in the strong coupling constant. The study is 
done for different jet clustering algorithms such as EM, JADE, E 
and DURHAM. Because three-jet observables are very sensitive
to the quark mass we consider the possibility of extracting 
the bottom quark mass from LEP data.
\end{quote}
\vspace*{\fill}
\begin{flushleft}
     FTUV/96-44 \\ IFIC/96-52 \\ \today
\end{flushleft}
\end{titlepage}

\newpage\addtocounter{footnote}{-1}


\begin{abstract}
We present the calculation of the decay width of the $Z$-boson 
into three jets including complete quark mass effects to 
second order in the strong coupling constant. The study is 
done for different jet clustering algorithms such as EM, JADE, E 
and DURHAM. Because three-jet observables are very sensitive
to the quark mass we consider the possibility of extracting 
the bottom quark mass from LEP data.
\end{abstract}

\maketitle

\def\mafigura#1#2#3#4{
  \begin{figure}[hbtp]
    \begin{center}
      \epsfxsize=#1
      \leavevmode
      \epsffile{#2}
    \end{center}
    \caption{#3}
    \label{#4}
  \end{figure} }

\section{INTRODUCTION}

A precise theoretical framework is needed for 
the study of the quark mass effects in 
physical observables because quarks are not free particles.
In fact, the quark masses should be seen more
like coupling constants than like physical parameters.
The perturbative pole mass and the running mass are
the two most commonly used quark mass definitions.
The perturbative pole mass, $M(p^2=M^2)$, 
is defined as the pole of the
renormalized quark propagator in a strictly perturbative sense.
It is gauge invariant and scheme independent. However, 
it appears to be ambiguous due to non-perturbative renormalons.
The running mass, $\bar{m}(\mu)$, the
renormalized mass in the $\overline{MS}$ scheme, does not suffer 
from this ambiguity.
Both quark mass definitions can be related perturbatively through
\beq
M = \bar{m}(\mu) \left\{ 1 + \frac{\as(\mu)}{\pi} \left[
\frac{4}{3}-\log \frac{\bar{m}^2(\mu)}{\mu^2} \right] \right\}.
\label{relates}
\eeq

Heavy quark masses, like the bottom quark mass, can be
extracted using QCD Sum Rules or lattice calculations
from the quarkonia spectrum, see \cite{UIMP}
and references therein. The bottom quark perturbative pole mass
appears to be around $M_b = 4.6-4.7 (GeV)$ whereas
the running mass at the running mass scale reads
$\bar{m}_b(\bar{m}_b) = (4.33 \pm 0.06) GeV$.
Performing the running until the $Z$-boson mass scale
we find $\bar{m}_b(M_Z) = (3.00 \pm 0.12) GeV$.
Since for the bottom quark the difference between the
perturbative pole mass and the
running mass at the $M_Z$ scale 
is quite significant it is crucial 
to specify in any theoretical perturbative
prediction at $M_Z$ which mass should we use.

The relative uncertainty in the strong coupling
constant decreases in the running from low to 
high energies as the ratio of the strong coupling constants
at both scales. On the contrary,
if we perform the quark mass running with the 
extreme mass and strong coupling constant values
and take the maximum difference as the propagated error,
induced by the strong coupling constant error,
the quark mass uncertainty increases following
\beq
\varepsilon_r (\bar{m}(M_Z)) \simeq 
\varepsilon_r (\bar{m}(\mu)) 
\eeq
\[ + \frac{2 \gamma_0}{\beta_0}
\left(\frac{\as(\mu)}{\as(M_Z)}-1 \right)
\varepsilon_r (\as(M_Z)),
\]
where $\gamma_0=2$ and $\beta_0=11-2/3 N_F$, see
figure~\ref{running}. We use the world average \cite{bethke} value 
$\as = 0.118 \pm 0.006$ for the strong coupling constant.

It is interesting to stress, looking at figure~\ref{running},
that even a big uncertainty in 
a possible evaluation of the bottom quark mass
at the $M_Z$ scale can be competitive with
low energy QCD Sum Rules and lattice calculations
with smaller errors~\footnote{A recent lattice
evaluation \cite{lattice} 
has enlarged the initial estimated error on the bottom quark 
mass, $\bar{m}_b(\bar{m}_b) = (4.15 \pm 0.20) GeV$, due to 
unknown higher orders in the perturbative matching of the HQET
to the full theory.}.
Furthermore, non-perturbative contributions are expected to
be negligible at the $Z$-boson mass scale.

The running mass holds another remarkable feature.
Total cross sections can exhibit potentially dangerous terms
of the type $M^2 \log M^2/s$ that however can be absorbed
\cite{chety} using Eq.~(\ref{relates}) and expressing the
total result in terms of the running mass.

\mafigura{5.5cm}{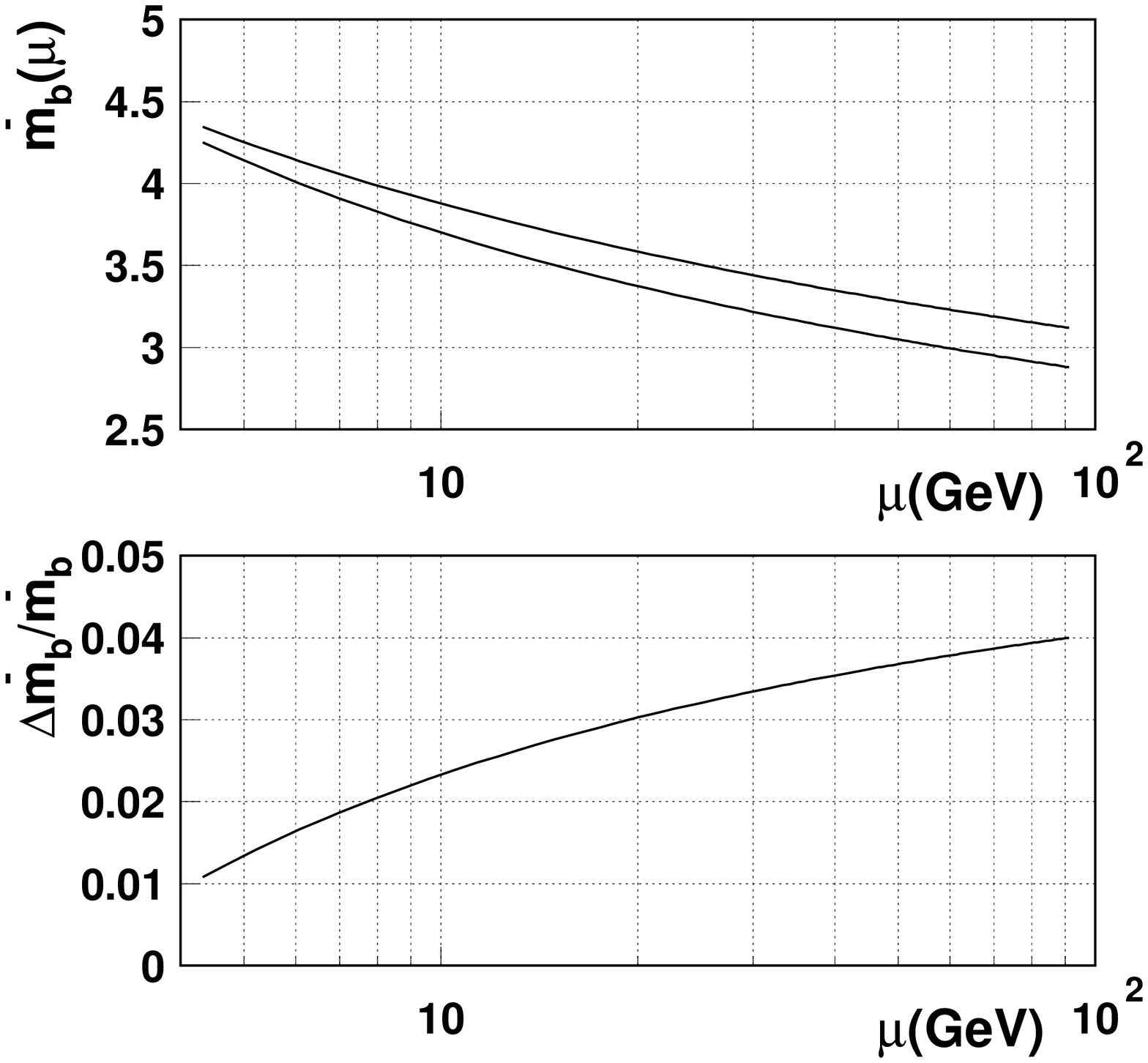}
{Running of the bottom quark mass from low energies to the
$M_Z$ scale.
Upper line is the run of $\bar{m}_b(\bar{m}_b)=4.39(GeV)$ with 
$\as (M_Z)=0.112$. Bottom line is the run of  
$\bar{m}_b(\bar{m}_b)=4.27(GeV)$ with $\as (M_Z)=0.124$.
Second picture is the difference of both, our estimate for
the propagated error.}
{running}

\section{THREE JETS OBSERVABLES AT LO}

Quark masses can be neglected for many observables
at LEP because usually they appear as the ratio
$m_q^2/M_Z^2$. For the heaviest quark produced at LEP,
the bottom quark, this means a correction of 3 per mil
for a quark mass of 5 (GeV).
Even if the coefficient in front is 10 we get at most a
3\% effect, 1\% if we use the bottom quark running mass
at $M_Z$.
This argument is true for total cross section. However,
jet cross sections depend on a new variable, $y_c$,
the jet-resolution parameter that defines the jet
multiplicity. This new variable introduces a new scale
in the analysis, $E_c = M_Z \sqrt{y_c}$, that
for small values of $y_c$ could enhance the effect
of the quark mass as $m_b^2/E_c^2 = (m_b^2/M_Z^2)/y_c$.
The high precision achieved at LEP makes these effects
relevant. In particular, it has been shown \cite{Juano}
that the biggest systematic error in the measurement of
$\as(M_Z)$ from $b\bar{b}$-production at LEP from the 
ratio of three to two jets comes from the uncertainties
in the estimate of the quark mass effects.

We are going to study the effect of the bottom quark mass 
in the following ratios of three-jet decay rates and 
angular distributions

\bea
R_3^{bd} &\equiv& \frac{\Gamma^b_{3j}(y_c)/\Gamma^b}
     {\Gamma^d_{3j}(y_c)/\Gamma^d}, \label{r3bd}\\
R^{bd}_\vartheta &\equiv& \left.
\frac{1}{\Gamma^b}\frac{d\Gamma^b_{3j}}{d\vartheta}\right/
\frac{1}{\Gamma^d}\frac{d\Gamma^d_{3j}}{d\vartheta},
\label{rtheta}
\eea
where we consider massless the $d$-quark and 
$\vartheta$ is the minimum of the angles formed between
the gluon jet and the quark and antiquark jets. Both 
observables are normalized to the total decay rates
in order to cancel large weak corrections dependent on the 
top quark mass \cite{top}.

\def\arraystretch{1.5}
\begin{table}[t]
\caption{The jet-clustering algorithms}
\label{algorithms}
\begin{center}
\begin{tabular}{ll}
\hline
Algorithm & Resolution \\
\hline
EM      & $2(p_i \cdot p_j)/s$ \\
JADE    & $2(E_i E_j)(1-\cos \vartheta_{ij})/s$ \\ 
E       & $(p_i+p_j)^2/s$ \\
DURHAM  & $2 \min(E_i^2,E_j^2)(1-\cos \vartheta_{ij})/s$  \\
\hline
\end{tabular}
\end{center}    
\end{table}

\mafigura{6cm}{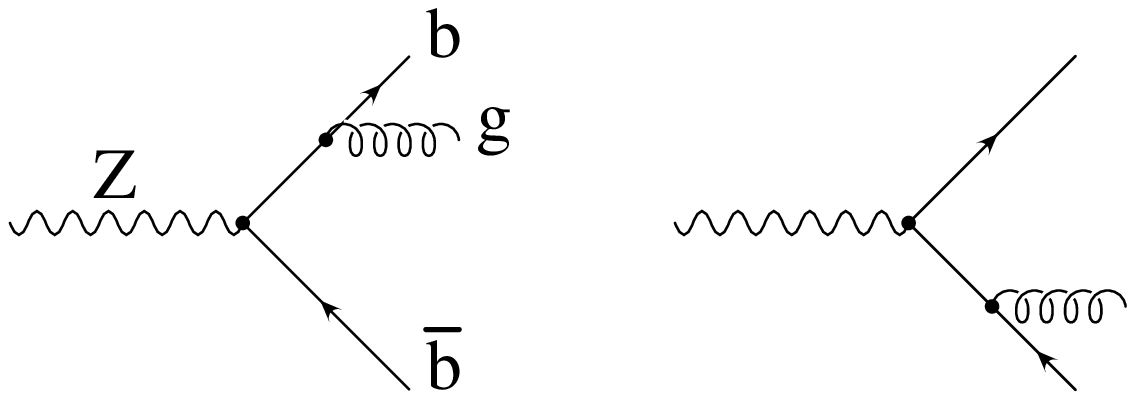}
{Feynman diagrams contributing to the three-jets decay rate
of $Z\rightarrow b\bar{b}$ at order $\as$.}
{feynmanLO}

At LO in the strong coupling constant
we must compute the amplitudes of the Feynman diagrams 
depicted in figure \ref{feynmanLO} plus the interchange of 
a virtual gluon between the quark and antiquark that 
only contributes to the two-jet decay rate.
In addition to renormalized UV divergences,
IR singularities, either collinear or soft,
appear because of the presence of massless
particles like gluons. Bloch-Nordsiek 
and Kinoshita-Lee-Nauemberg theorems\cite{IR} assure
IR divergences cancel for inclusive cross section.
Technically this means, if we use 
DR to regularize the IR divergences of the loop diagrams we 
should express the phase space for the tree-level diagrams
in arbitrary $D$-dimensions. The IR singularities 
cancel when we integrate over the full phase space.

Another delicate question is the problem of
hadronization. Perturbative QCD gives results at the level
of partons, quarks and gluons, but in nature one observes
hadrons, not partons, and hadronization can shift the QCD
predictions.
We apply to the parton amplitudes the same jet clustering
algorithms applied experimentally to the real observed particles,
see table~\ref{algorithms}.
Starting from a bunch of particles of momenta $p_i$ we calculate,
for instance, $y_{ij}=2(p_i \cdot p_j)/s$, the scalar product
of all the possible momenta pairs.
If the minimum is smaller than a fixed $y_c$
we combine the two involved particles in a new pseudoparticle
of momentum $p_i+p_j$. The procedure starts again until all
the $y_{ij}$ are bigger than $y_c$. The number of pseudoparticles
at the end of the procedure defines the number of jets.
The jet clustering algorithms
automatically define IR finite quantities.
For the moment, we do not enter in the question of which
is the best jet clustering algorithm
although the main criteria followed to choose
one of them should be based in two requirements: minimization
of higher order corrections and insensitivity to hadronization.
If we restrict to the three-jet decay rate the
IR problem can be overcome and everything can be calculated
in four dimensions because
the jet clustering algorithms automatically exclude 
the IR region from the three-body phase space.

For massless particles and at the lowest order the EM~\cite{LO},
JADE and E algorithms give the same answers.
Analytical results for the massless 
three-jet fraction exist for both JADE-like~\cite{KN} and 
DURHAM~\cite{Durham} algorithms.
A complete analysis for the ratios of three-jet decay 
rates and the angular distributions quoted in Eq.~\ref{r3bd} 
and \ref{rtheta} can be found in \cite{LO}.
For practical purposes a parametrization of the result in terms of
a power series in $\log y_c$ gives a good description \cite{KN,LO}.

\mafigura{6cm}{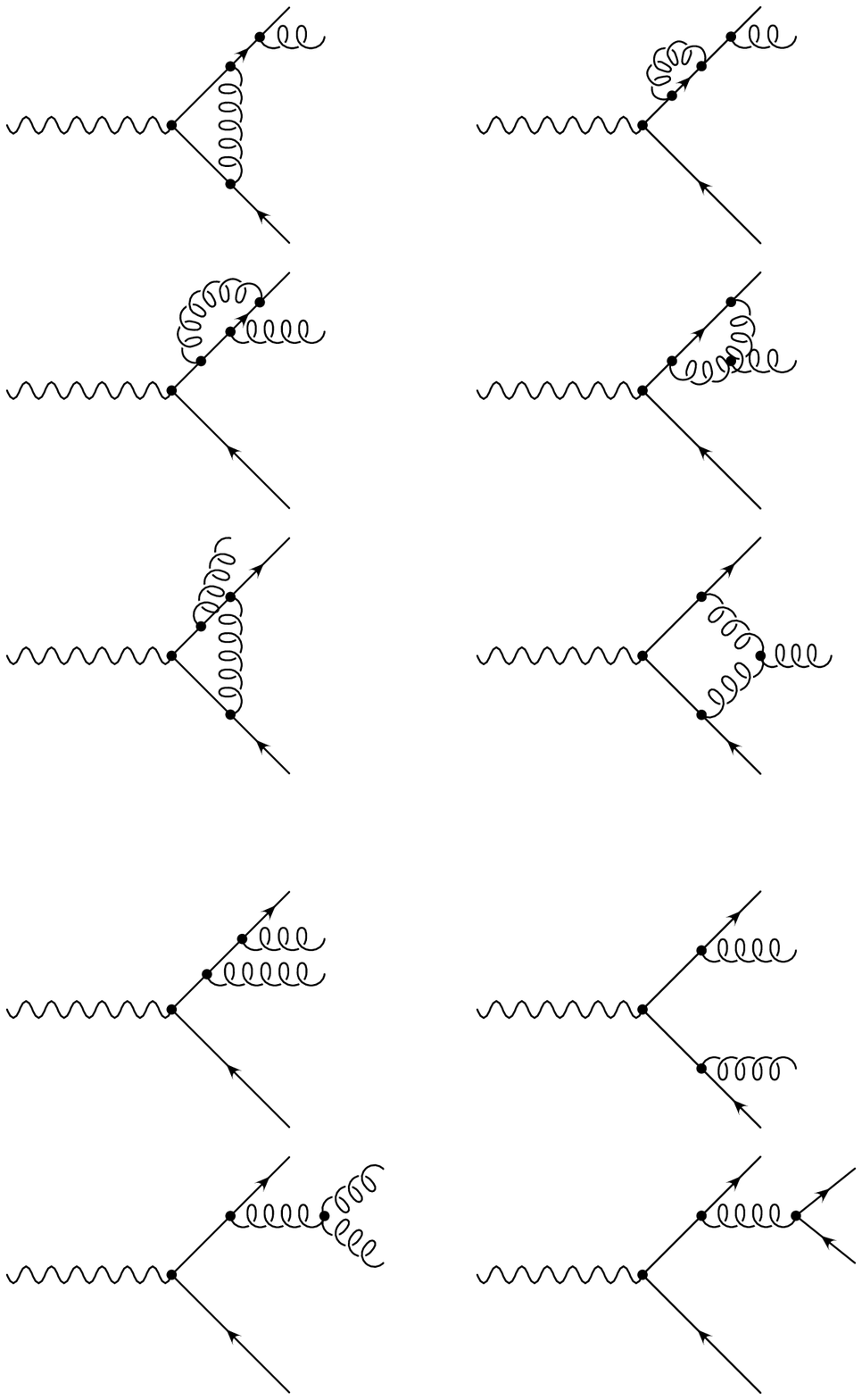}
{Feynman diagrams contributing to the three-jets decay rate
of $Z\rightarrow b\bar{b}$ at order $\as^2$.
Self-energies in external legs have not been shown.}
{feynmanNLO}

\section{THREE JETS OBSERVABLES AT NLO}

The effect of the bottom quark mass has been studied
experimentally by~\cite{Joan} on the $R_3^{bd}$ ratio.
As we have seen the running of
the bottom quark mass from low energies to the $M_Z$ scale is 
quite strong. The LO QCD prediction for 
$R_3^{bd}$ does not allow us to distinguish which mass we 
should use in the theoretical expressions, either the pole mass
or the running mass at some scale. The computation of the 
NLO is mandatory if we want to extract information about 
the bottom quark mass from LEP data.

At the NLO we have to calculate 
the interference of the loop diagrams depicted in
figure~\ref{feynmanNLO} with the lowest order Feynman diagrams
of figure~\ref{feynmanLO} plus the square of the tree-level 
diagrams of figure~\ref{feynmanNLO}.
The amplitudes in the massless case were
calculated by~\cite{ERT,KL}.
The implementation of the jet clustering algorithms 
was performed by~\cite{KN}.

The main problem that now we can not avoid is the appearance 
of IR singularities. With massive quarks 
we loose all the quark-gluon collinear divergences.
The amplitudes behave better in the IR region.
The disadvantage however is the mass itself.
We have to perform quite more complicated 
loop and phase space integrals. Furthermore, we still conserve the 
gluon-gluon collinear divergences leading to IR double poles.

The three-jet decay rate can be written as 
\beq
\Gamma^{b}_{3j} = C [g_V^2 H_V(y_c,r_b) + g_A^2 H_A(y_c,r_b)],
\eeq
where $r_b=m_b^2/M_Z^2$, 
$C=M_Z \: g^2/(c_W^2 64 \pi) (\as/\pi)$ is a normalization constant
that disappear in the ratio and
$g_V$ and $g_A$ are the vector and the axial-vector
neutral current quark couplings.
At tree-level and for the bottom quark 
$g_V = -1 + 4 s_W^2/3$ and $g_A = 1$.
Now we can expand the functions $H_{V(A)}$
in $\as$ and factorize the leading dependence on the quark mass 
as follows
\bea
H_{V(A)} &=& A^{(0)}(y_c) + r_b B_{V(A)}^{(0)}(y_c,r_b) \\
&+& \frac{\as}{\pi} \left( A^{(1)}(y_c) + r_b B_{V(A)}^{(1)}(y_c,r_b) 
\right), \nonumber
\eea
where we have taken into account that for massless quarks vector
and axial contributions are identical\footnote{We do not consider the
small $O(\as^2)$ triangle anomaly \cite{triangle}.
With our choice of the normalization $A^{(0)}(y_c)=A(y_c)/2$ and 
$A^{(1)}(y_c)=B(y_c)/4$, where $A(y_c)$ and $B(y_c)$ are 
defined in \cite{KN}.}.

First steep in the calculation is to show the cancellation of
the IR divergences in order to build matrix elements free of
singularities. It is possible to do it analytically.
However, we knew from the beginning IR divergences should 
disappear \cite{IR}. The challenge is in the calculation of the 
finite parts. This calculation is rather long, complex and full 
of difficulties. Strong cancellations occur between different
groups of diagrams making difficult even a numerical approach.
We have taken as guide line the massless result of \cite{ERT,KN}
although the IR structure of the massive case is completely
different from the massless one.

\mafigura{7cm}{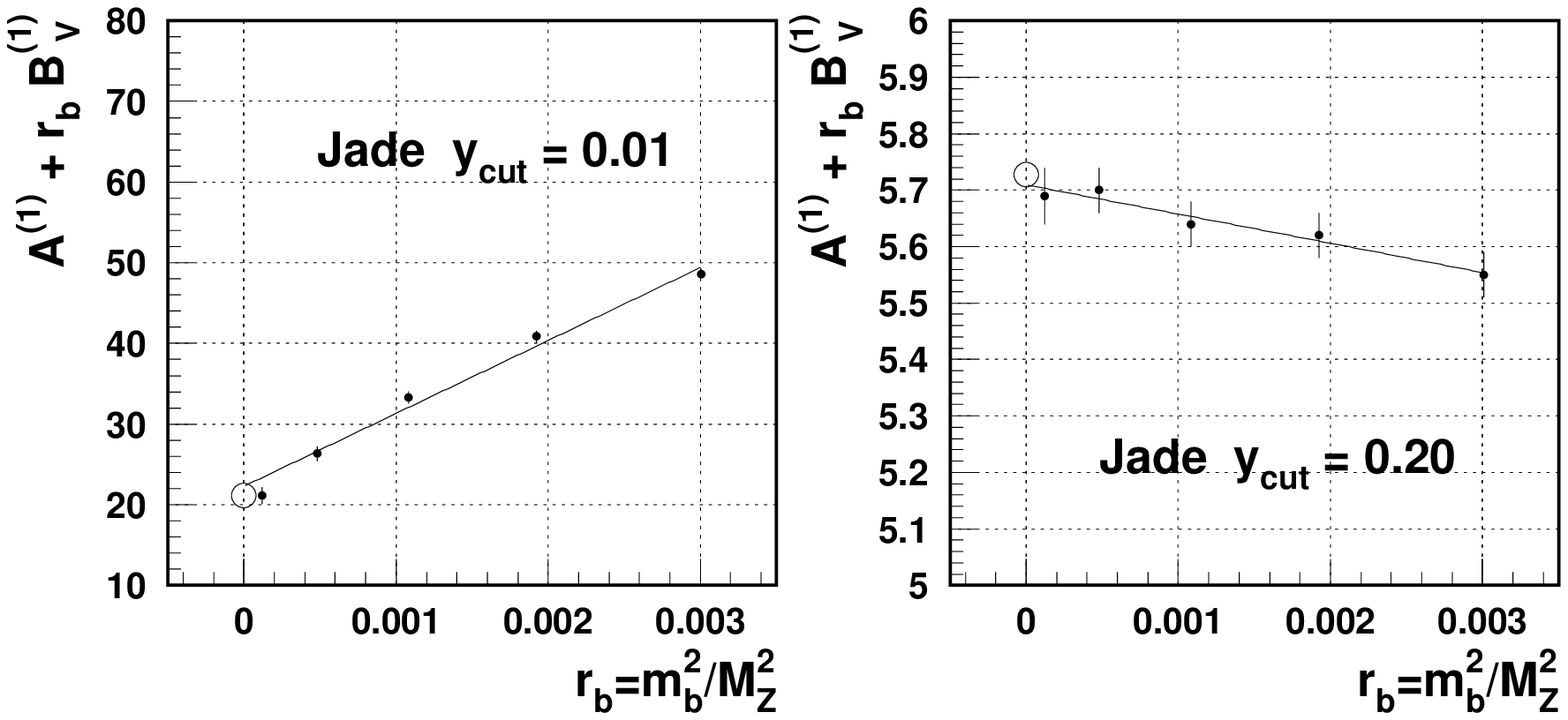}
{NLO vector contribution to the three-jet decay rate of 
$Z\rightarrow b\bar{b}$ for bottom quark masses from
$1$ to $5(GeV)$ and fixed $y_c$ in the JADE algorithm.
Big circle is the massless case.}
{Jtest}
\mafigura{7cm}{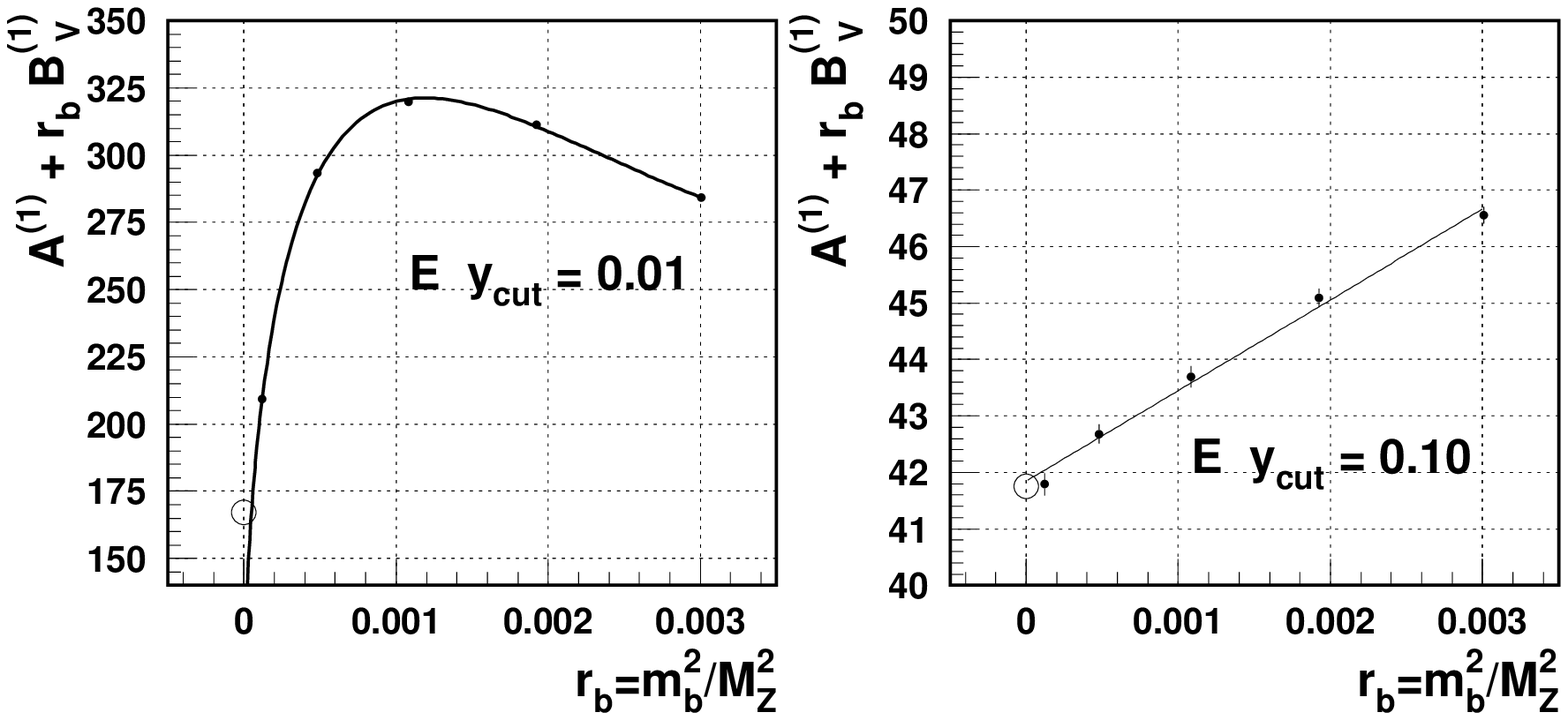}
{NLO vector contribution to the three-jet decay rate of
$Z\rightarrow b\bar{b}$ for bottom quark masses from 
$1$ to $5(GeV)$  and fixed $y_c$ in the E algorithm.
Big circle is the massless case.}
{Etest}

In figures \ref{Jtest} and \ref{Etest} we present our
preliminary result for the vectorial contribution to the
$O(\as^2)$ three-jet decay rate of the Z-boson into
bottom quarks. We have performed the calculation for
different values of the bottom quark mass from $1$ to $5(GeV)$
for fixed $y_c$.
We want to show we can recover the massless
result \cite{KN}, depicted as a big circle, i.e., in the limit
of massless quarks we reach the $A^{(1)}(y_c)$ function.
This is our main test to have confidence in our calculation.

In the JADE algorithm we can see that for big values of
$y_c$ the NLO corrections due to the quark mass are
very small and below the massless result.
Notice they increase quite a lot for small values of $y_c$
and give a positive correction that will produce a change 
in the slope of the LO prediction for $R_3^{bd}$.
In any case we recover
the massless limit and a linear
parametrization in the quark mass squared could
provide a good description. 

The E algorithm behaves also linearly in the quark mass squared
although only for big values of $y_c$.
Corrections in the E algorithm are always very
strong. The reason is the following, the resolution parameter
for the E algorithm explicitly incorporates the quark mass,
$y_{ij}=(p_i+p_j)/s$, i.e., for the same value of $y_c$
we are closer to the two-jet IR region and the difference
from the other algorithms is precisely the quark mass.
This phenomenon already manifest
at the LO. The behaviour of the E algorithm is completely
different from the others for massive quarks.
It is difficult to believe in the E algorithm as a
good prescription for physical applications 
since mass corrections as so big. However for the same reason,
it seems to be the best one for testing massive calculations.

\section{CONCLUSIONS}
We have presented the first results for the NLO strong 
corrections to the three-jet decay rate of the Z-boson into
massive quarks. In particular, extrapolating our result
we have shown we can recover previous calculations with
massless quarks.
Their application to LEP data,
together with the already known LO, 
can provide a new 
way for determining the bottom quark mass
and to show for the first time its running.

\vskip 5mm
\noindent{\bf Acknowledgements.} 
I would like to thank
J.~Fuster for very encouraging comments during the 
development of this calculation and for carefully 
reading this manuscript,  A.~Santamaria 
for very useful discussions and 
S. Narison for the very kind atmosphere created 
at Montpellier.


\end{document}